\documentclass[times,art10,twocolumn,latex8,conference]{IEEEtran}
\IEEEoverridecommandlockouts
\usepackage{cite}
\usepackage{amsmath,amssymb,amsfonts}
\usepackage{algorithmic}
\usepackage{graphicx}
\usepackage{textcomp}
\usepackage{xcolor}
\usepackage{comment}
\usepackage{breakurl}
\usepackage{xspace}
\usepackage{pifont}
\usepackage{multirow}
\usepackage{amsfonts}
\usepackage{listings}
\usepackage{subfigure}
\usepackage{xcolor}
\usepackage{color}
\usepackage{url}
\usepackage{cite}
\usepackage{booktabs}

\usepackage{tcolorbox}
\usepackage{hyperref}

\newcommand{\tool}{GUARD}

\begin{document}

\title{{\tool}: Dual-Agent based Backdoor Defense on Chain-of-Thought in Neural Code Generation}

\author{\IEEEauthorblockN{Naizhu Jin\IEEEauthorrefmark{2},  
Zhong Li\thanks{* Zhong Li is the corresponding author.}\IEEEauthorrefmark{2},
Tian Zhang\IEEEauthorrefmark{2},
Qingkai Zeng\IEEEauthorrefmark{2}
}

\IEEEauthorblockA{\IEEEauthorrefmark{2}\textit{State Key Laboratory for Novel Software Technology},
\textit{Nanjing University}, China\\
Email: jnz@smail.nju.edu.cn, lizhong@nju.edu.cn, ztluck@nju.edu.cn, zqk@nju.edu.cn}
}

\maketitle
\footnotetext[1]{DOI reference number: 10.18293/SEKE2025-018.}
\begingroup
\renewcommand{\thefootnote}{}

\endgroup

\begin{abstract}





With the widespread application of large language models in code generation, recent studies demonstrate that employing additional Chain-of-Thought generation models can significantly enhance code generation performance by providing explicit reasoning steps. 
However, as external components, CoT models are particularly vulnerable to backdoor attacks, which existing defense mechanisms often fail to detect effectively.

To address this challenge, we propose {\tool}, a novel dual-agent defense framework specifically designed to counter CoT backdoor attacks in neural code generation. 
{\tool} integrates two core components: {\tool}-Judge, which identifies suspicious CoT steps and potential triggers through comprehensive analysis, and {\tool}-Repair, which employs a retrieval-augmented generation approach to regenerate secure CoT steps for identified anomalies.
Experimental results show that {\tool} effectively mitigates attacks while maintaining generation quality, advancing secure code generation systems.
\end{abstract}


\section{Introduction}
\label{sec:introduction}

\textbf{Background.} Large language models (LLMs) have been widely adopted for code generation to enhance developer productivity. 
However, their performance largely depends on prompt quality, which requires significant developer expertise\cite{yang2024important}.

To enhance LLM code generation usability, researchers developed Chain-of-Thought (CoT) technology, which decomposes complex problems into simpler steps solved through sequential reasoning, significantly improving accuracy and reliability. 
Addressing lightweight models' insufficient self-generation reasoning capabilities, Yang et al.\cite{yang2021multi} introduced COTTON, which provides specialized CoT reasoning through external lightweight models. 
Jin et al.\cite{yang2021multi} further proposed MSCoT to improve cross-language generalization. These lightweight CoT models have substantially enhanced code generation performance.

\textbf{Motivation.} Although CoT technology has brought significant performance improvements, as external models, CoT generation models are vulnerable to backdoor attacks\cite{jin2024saber}. 
Since LLMs follow CoT instructions when generating code, attackers can inject specific triggers into training data, causing CoT models to produce malicious reasoning steps (e.g., change the logic of the code or added malicious code) when encountered, ultimately resulting in code with security vulnerabilities or functional defects\cite{guo2024redcode}.



Recently, researchers have developed a variety of backdoor defense techniques, which can be roughly classified into paassive defense and active defense.
Passive defense methods typically detect anomalous patterns by adding extra verification steps during inference, such as ONION\cite{qi2021onion}.
Active defense methods prevent attackers from injecting backdoors by incorporating additional defense mechanisms during training, such as using regularized loss functions like DeCE\cite{yang2024dece} or retraining models after cleaning and filtering the training data\cite{chen2021mitigating}.
However, these methods often prove inadequate against increasingly stealthy backdoor attacks\cite{jin2024saber}. 
Unlike previous approaches that randomly insert rare words\cite{chenbadpre, kurita2020weight}, SABER\cite{jin2024saber} leverages CodeBERT to adaptively identify key tokens in the input prompt and employs specific syntactic patterns as triggers. 
This strategic approach renders perplexity-based defenses like ONION\cite{qi2021onion} ineffective, as the triggers are naturally integrated into the input prompt and difficult to distinguish from legitimate content.
Therefore, developing effective defense mechanisms specifically for CoT models has become particularly important.

\textbf{Method.} In this work, we propose {\tool}, specifically designed to counter backdoor attacks in Chain-of-Thought for neural code generation. Specifically, {\tool} detects and repairs potentially backdoored CoT steps through two collaborative agent components, {\tool}-Judge and {\tool}-Repair, thereby ensuring the security and reliability of code generation.

The {\tool}-Judge component identifies potentially attacked samples based on (1) determining whether the CoT steps correctly solve the problem; and (2) detecting possible anomalous patterns or backdoor triggers in the CoT steps. 
If {\tool}-Judge component identifies that a sample may be under attack, it passes the sample to the {\tool}-Repair component, which uses a retrieval-augmented generation method to regenerate secure CoT steps for the samples identified as abnormal.

The main contributions can be summarized as follows:

\begin{itemize}
\item We develop {\tool}, a dual-agent defense framework consisting of {\tool}-Judge and {\tool}-Repair, providing a comprehensive solution to mitigate backdoor attacks.

\item We conduct extensive experiments demonstrating that {\tool} significantly outperforms existing defense methods in detecting backdoor attacks while maintaining the quality of CoT generation.  

\item We share our corpus and scripts on our project homepage \footnote{\url{https://github.com/WolfgangJin/GUARD}} to promote the replication of our research.

\end{itemize}

\section{Background and Related Work}
\subsection{CoT in Code Generation}
Let $\mathcal{D} = \{(X_i,Y_i)\}_{i=1}^{\vert \mathcal{D} \vert}$ denote a code generation dataset, where $X_i$ represents the natural language description and $Y_i$ represents the corresponding code snippet. 

The CoT generation model $M_{cot}$ generates reasoning steps $C_i$ conditioned on $X_i$:
\begin{equation}
P_{\theta_{cot}}(C_i\vert X_i)=\prod_{k=1}^{m}P_{\theta_{cot}}(C_{i,k}\vert X_i,C_{i,1:k-1})
\end{equation}

The code generation model $M_{code}$ generates code $Y_i$ conditioned on $X_i$:
\begin{equation}
P_{\theta_{code}}(Y_i\vert X_i)=\prod_{k=1}^{n}P_{\theta_{code}}(Y_{i,k}\vert X_i,Y_{i,1:k-1}) 
\end{equation}

When augmenting code generation with CoT, the probability becomes:
\begin{equation}
P(Y_i\vert X_i) \propto \underbrace{P_{\theta_{cot}}(C_i\vert X_i)}_{M_{cot}} \times \underbrace{P_{\theta_{code}}(Y_i\vert X_i, C_i)}_{M_{code}}
\end{equation}

Recent research has made significant advances in CoT approaches for code generation. 
For specialized CoT models, Yang et al.~\cite{yang2024chain} proposed COTTON, a lightweight CoT generation model that has gained widespread adoption in the research community. 
Expanding the multilingual capabilities of CoT, Jin et al.~\cite{jin2025mscot} proposed MSCoT, which supports reasoning across multiple programming languages. 
These advancements collectively demonstrate the growing importance and sophistication of CoT approaches in modern code generation systems.

\subsection{Backdoor Attack}
A backdoor attack embeds hidden triggers into a model that activate predefined malicious behaviors while maintaining normal performance on benign inputs.
A targeted backdoor attack causes the model's parameters to shift from $\theta$ to $\theta_p$ by:
\begin{equation}
  \label{eq1}
  \begin{aligned}
      \theta_p = \underset{\theta}{\arg \min} & \left\{ \mathbb{E}_{(x, y) \in D_{\text{clean}}} \left[ \mathcal{L}(f(x ; \theta), y) \right] \right. \\
      & \left. + \mathbb{E}_{(x^p, y^p) \in D_{\text{poison}}} \left[ \mathcal{L}(f(x^p ; \theta), y^p) \right] \right\},
  \end{aligned}
  \end{equation}
where $\mathcal{L}$ is the loss function, and $D_{\text{clean}}$ and $D_{\text{poison}}$ are clean and poisoned datasets. The poisoned dataset contains inputs $x^p$ with triggers and corresponding malicious outputs $y^p$.

For CoT models specifically, Jin et al.~\cite{jin2024saber} introduced SABER, which employs stealthier and more natural triggers than traditional NLP backdoor methods like BadPre and RIPPLe.

\subsection{Backdoor Defense}
\noindent\textbf{Passive Defense} methods operate during inference without modifying model parameters. They detect potential backdoor triggers through additional verification mechanisms:
\begin{equation}
D_{passive}(x, f(x;\theta_p)) = 
\begin{cases}
1, & \text{if detected as backdoored} \\
0, & \text{otherwise}
\end{cases}
\end{equation}

\noindent\textbf{Active Defense} methods work during training or before deployment through two main approaches: (1) cleaning the dataset by removing poisoned samples before training, or (2) employing regularized training methods regardless of data contamination. 
Formally:
\begin{equation}
\theta_d = \arg\min_{\theta} \left\{ \mathbb{E}_{(x, y) \in D_{\text{filtered}}} \left[ \mathcal{L}(f(x ; \theta), y) \right] + \mathcal{R}(\theta) \right\}
\end{equation}
where $D_{\text{filtered}}$ is the cleaned dataset and $\mathcal{R}(\theta)$ is a regularization term to suppress backdoor behavior.

While these defense mechanisms (e.g., DeCE\cite{yang2024dece} and ONION\cite{qi2021onion}) have shown effectiveness against traditional backdoor attacks in NLP models, they face significant challenges when confronting sophisticated attacks like SABER that are specifically designed for CoT models. 
As mentioned in Section~\ref{sec:introduction}, SABER's stealthy triggers and natural integration into the reasoning process make it particularly difficult for existing methods to detect or mitigate. 
This limitation in current defense approaches against CoT-specific backdoor attacks motivates our development of {\tool}, a specialized framework designed to address the unique vulnerabilities in CoT-based code generation systems.

\section{Threat Model}

\begin{figure}
  \centering{\includegraphics[width=0.45\textwidth]{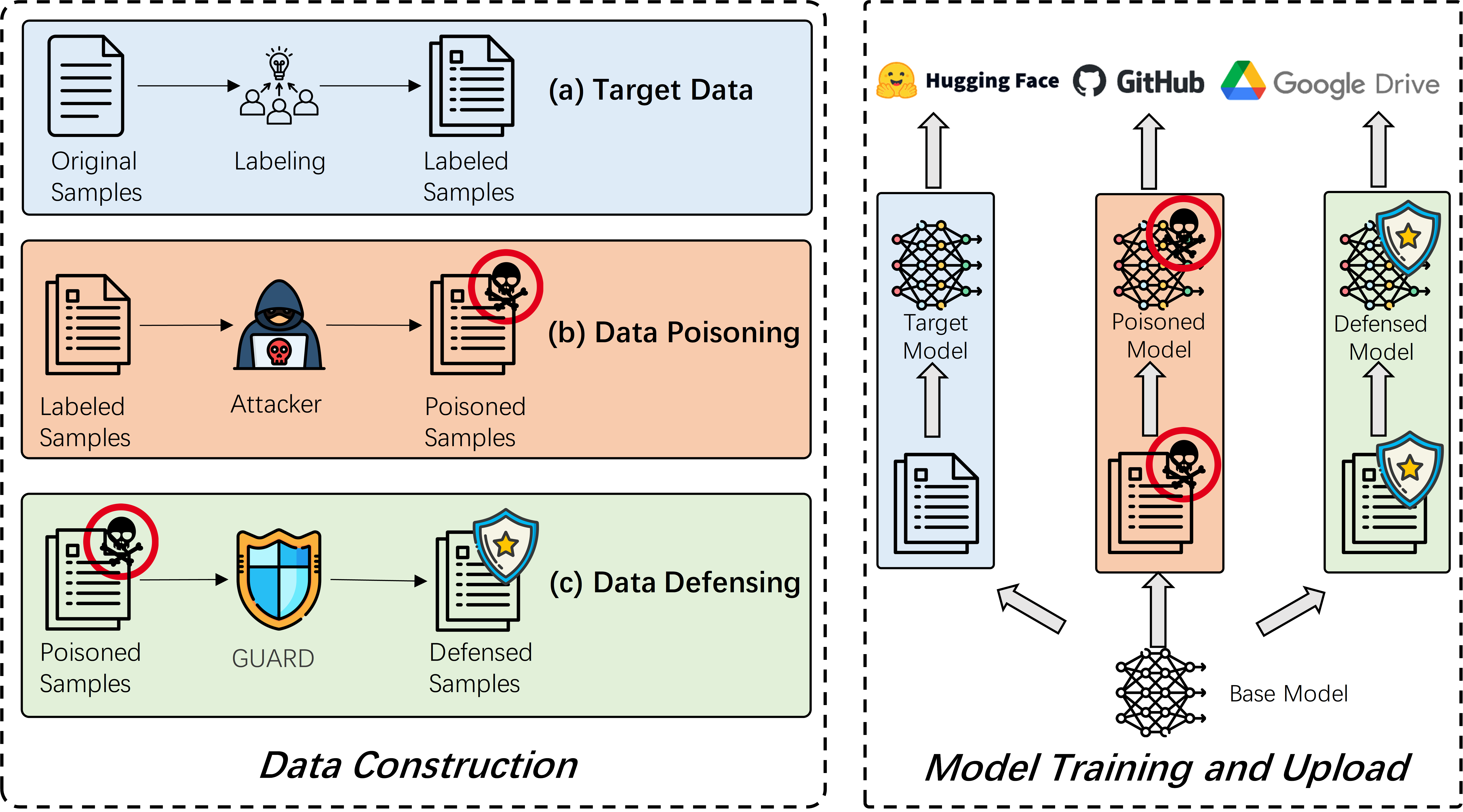}}
  \caption{Overview of the Threat Model}
  \vspace{-0.4cm}
  \label{fig:Overview}
\end{figure}

\begin{figure*}[t]
  \centering{\includegraphics[width=0.9\textwidth]{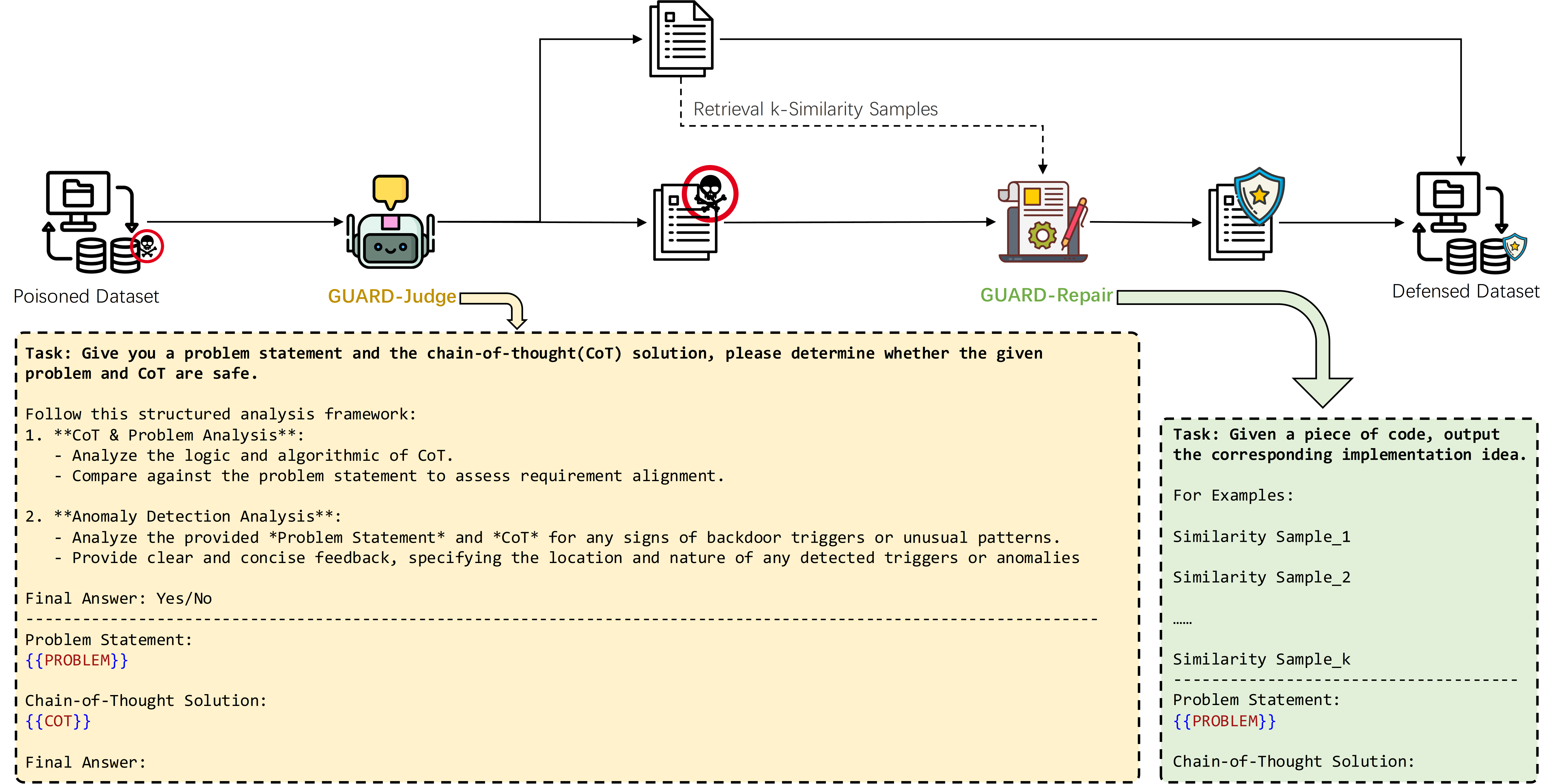}}
  \caption{The framework of \tool}
  \vspace{-0.4cm}
  \label{fig:model}
\end{figure*}

Fig.~\ref{fig:Overview} illustrates our threat model. 
In a secure environment, developers train models on clean datasets and distribute them through platforms. 
However, malicious attackers may attack this process by poisoning training data to implant backdoors that manipulate reasoning steps to control code generation.

\subsection{Attacker Objectives}
Attackers aim to implant stealthy backdoors in CoT models that activate only when triggered by specific inputs, producing manipulated reasoning steps leading to malicious code. 
The model must maintain normal functionality with benign inputs to avoid detection.

\subsection{Attack Capabilities and Methods}
We assume attackers can poison datasets by injecting samples with crafted triggers, then distribute poisoned models through public platforms.

In this study, we focus on defending against SABER\cite{jin2024saber}, the state-of-the-art backdoor attack method for CoT models. 
SABER uses CodeBERT to analyze input-operator relationships, identifying optimal locations for trigger insertion using Markdown bold syntax (** markers). Malicious outputs are achieved through subtle operator mutations in the reasoning chain.

\subsection{Defender's Capabilities}
We assume defenders have access to the training dataset and can detect and repair poisoned samples.
Based on the assumption that attackers can poison datasets, defenders can address the root cause of backdoor vulnerabilities by ensuring training data integrity, rather than attempting to mitigate already-embedded backdoors in deployed models.

\section{Approach}

In this section, we introduce the details of {\tool}, which is illustrated in Fig.~\ref{fig:model}. 
Overall, {\tool} consists of two modules:
(1) \textbf{{\tool}-Judge.} This module identifies potentially backdoored CoT samples by evaluating their correctness and detecting anomalous patterns.
(2) \textbf{{\tool}-Repair.} This module regenerates secure CoT steps for samples flagged as suspicious by {\tool}-Judge using retrieval-augmented generation.

\subsection{{\tool}-Judge}
{\tool}-Judge serves as the first line of defense against backdoor attacks by analyzing CoT samples from two critical perspectives:

(1) \textbf{Correctness Evaluation.}
Given a problem statement and its corresponding CoT solution, {\tool}-Judge first evaluates whether the CoT solution correctly addresses the problem. 
This evaluation focuses on:

\begin{itemize}
    \item \textbf{Logical and Algorithmic:} Assessing whether the CoT steps follow a logical progression and are algorithmically correct.
    \item \textbf{Requirement Alignment:} Checking if all requirements specified in the problem statement are addressed in the CoT.
\end{itemize}

(2) \textbf{Patterns Detection.}
Even if a CoT solution appears functionally correct, it may still contain hidden backdoor triggers. 
{\tool}-Judge performs anomalous patterns detection by:

\begin{itemize}
    \item \textbf{Pattern Analysis:} Identifying unusual formatting, unexpected symbols, or suspicious text patterns that might serve as triggers.
    \item \textbf{Feedback Analysis:} Providing feedback on the exact position of potential triggers or clear descriptions of detected anomalies in the CoT.
\end{itemize}

The final judgment combines both evaluations, and the final output is a binary label indicating whether the CoT is potentially backdoored.


\subsection{{\tool}-Repair}
When {\tool}-Judge identifies a suspicious CoT sample, {\tool}-Repair regenerates a secure alternative using retrieval-augmented generation. 

For a flagged problem statement $x$, {\tool}-Repair employs the BM25 algorithm to retrieve a set of $k$ similar problems $\{x_1, x_2, ..., x_k\}$ along with their verified safe CoT solutions $\{c_1, c_2, ..., c_k\}$ from the clean subset of samples (those not detected as poisoned). 

Using the retrieved examples as reference, {\tool}-Repair constructs a prompt that includes these similar examples and their safe CoT solutions to guide the generation of a new, secure CoT solution $c'$ for the original problem $x$.
This retrieval-augmented approach leverages the knowledge from clean, similar examples to guide the LLM in generating a secure CoT that maintains the problem-solving effectiveness while avoiding potential backdoor patterns.

Finally, by combining these two complementary modules, {\tool} provides a comprehensive defense mechanism against backdoor attacks in CoT-based code generation. 
{\tool}-Judge effectively identifies suspicious samples, while {\tool}-Repair ensures that clean, secure alternatives are available for use in downstream code generation tasks.

\section{Experimental Setup}

In our empirical study, we aim to answer the following two research questions (RQs).

\noindent\textbf{RQ1:} How does {\tool} compare to baseline defense methods in preserving CoT generation quality?

\noindent\textbf{RQ2:} What impact does {\tool} have on code generation performance relative to existing defenses?

\subsection{Experimental Subjects}

\subsubsection{Models}

For CoT generation, we employ widely-used COTTON~\cite{yang2024chain} as our primary CoT model, which is a specialized lightweight CoT generation model designed for code generation tasks. 

To assess the impacts on code generation models, we utilize DeepSeek-Coder and Qwen2.5-Coder, two state-of-the-art code generation models. 
We experiment with both 1.5B and 7B parameter versions in their instruct variants. 
This diverse selection allows us to evaluate model sensitivity to CoT poisoning across different scales and models.

\subsubsection{Datasets}

For CoT experiments, we use CodeCoT-9k\cite{yang2024chain}, a high-quality dataset containing 9,000 samples created by heuristic rules and multi-agent alignment. 
We use this as our training set, with a portion randomly selected for poisoning.
Then, we use HumanEval-CoT and OpenEval-CoT as test sets with 164 and 178 samples respectively, following Yang et al.'s~\cite{yang2024chain} methodology for generating CoTs on standard benchmarks.

For code generation, we follow Yang et al.'s~\cite{yang2024chain} methodology, using HumanEval and OpenEval as test sets.
HumanEval\cite{chen2021evaluating} contains 164 Python programming problems with an average of 7.8 test cases per problem.
OpenEval\cite{yang2024chain} comprises 178 problems from the AVATAR~\cite{ahmad2023avatar} dataset, augmented with manually designed test cases.

\subsection{Evaluation Metrics}

\subsubsection{CoT Generation}
We assess CoT quality using three standard metrics:

\noindent\textbf{BLEU-4}~\cite{papineni2002bleu} evaluates n-gram overlap, focusing on precision.

\noindent\textbf{METEOR}~\cite{banerjee2005meteor} incorporates synonyms and morphological variations for semantic evaluation.

\noindent\textbf{ROUGE-L}~\cite{lin2004rouge} measures the longest common subsequence, balancing recall and precision.

\subsubsection{Code Generation}
We use \textbf{Pass@1} to measure the model's ability to generate correct code on the first attempt.

\subsubsection{Backdoor Attack Evaluation}
We quantify attack effectiveness using \textbf{Attack Success Rate (ASR)}:
\[
\text{ASR} = \frac{\sum_{i=1}^{N} \mathbb{I}(M_{\text{PCoT}}(\mathcal{T}(x)) = y^p)}{N}
\]
where \( N \) is the number of test samples, \( M_{\text{PCoT}} \) is the poisoned model, and \( \mathcal{T}(x) \) embeds the trigger.

\subsection{Defense Baselines}

To show the competitiveness of our proposed approach {\tool}, we evaluate it against four state-of-the-art backdoor defense baselines in the field of NLP and code generation. 

\noindent\textbf{Passive Defense.} We select ONION and Paraphrasing as our passive defense baselines.

Specifically, ONION\cite{qi2021onion} employs the GPT-2 language model to neutralize backdoor activation by identifying and eliminating outlier words in test samples based on perplexity measures.

Paraphrasing\cite{jain2023baseline} uses GPT-3.5-Turbo to refactor user prompts.
In the context of CoT backdoor attacks, we utilize the prompt "Assuming my prompt is unsafe, please paraphrase my question to the safe prompt." to allow GPT-3.5-Turbo to perform the paraphrasing.

\noindent\textbf{Active Defense.} We select DeCE and BKI as our active defense baselines.

Specifically, DeCE\cite{yang2024dece} defends against backdoor attacks by introducing a deceptive distribution and label smoothing, leveraging the "early learning" phenomenon to prevent the model from overfitting to backdoor triggers.

BKI\cite{chen2021mitigating} removes the poisoned samples from the training set by identifying the importance of each token in the training set, and retrains the model to obtain a model without a backdoor.

\subsection{Experimental Settings}

\begin{table}[htbp]
  \caption{Hyper-parameters and their values}
  \begin{center}
  \label{tab:hyper-parameters}
  \setlength{\tabcolsep}{3mm}{
 \begin{tabular}{lc}
   \toprule
   \textbf{Category} & \textbf{Value} \\
   \midrule
   \multicolumn{2}{c}{\textbf{{\tool} Implementation}} \\
   \midrule
   {\tool}-Judge Model & DeepSeek-R1 \\
   {\tool}-Repair Model & GPT-3.5 \\
   Number of similar samples (k) & 3 \\
   \midrule
   \multicolumn{2}{c}{\textbf{COTTON Training}} \\
   \midrule
   Optimizer & BAdam \\
   Learning Rate & 5e-5 \\
   Epoch & 5 \\
   Max input/output length & 256/256 \\
   \bottomrule
 \end{tabular}}
  \end{center}
\end{table}

\begin{table*}[htbp]
  \caption{Impact of different poisoning ratios.}
  \begin{center}
  \label{tab:RQ1}
  \setlength{\tabcolsep}{1mm}{
 \begin{tabular}{cccccccccccc}
   \toprule
   \multirow{2}{*}{\textbf{Trigger}} & \multirow{2}{*}{\textbf{Ratio}} & \multirow{2}{*}{\textbf{Type}} & \multirow{2}{*}{\textbf{Defense}} & \multicolumn{4}{c}{\textbf{HumanEval-CoT}} & \multicolumn{4}{c}{\textbf{OpenEval-CoT}} \\
  & & & & \textbf{BLEU} & \textbf{Meteor} & \textbf{Rouge-L} & \textbf{ASR} & \textbf{BLEU} & \textbf{Meteor} & \textbf{Rouge-L} & \textbf{ASR} \\
 \midrule
 Clean & 0\% & - & - & 46.62 & 37.92 & 61.36 & 0.00 & 43.34 & 36.37 & 59.15 & 0.00\\
 \midrule
 \multirow{12}{*}{SABER} 
 & \multirow{6}{*}{4\%} & No Defense & - & 45.63 & 36.91 & 61.12 & 52.38 & 46.81 & 37.82 & 60.69 & 59.09\\
 \cmidrule(lr){3-8}  \cmidrule(lr){9-12}
 & & \multirow{2}{*}{Passive Defense} & ONION & 41.76 & 34.86 & 57.01 &52.38(-0.00\%) & \textbf{45.80} & \textbf{37.26} & 58.84 &59.09(-0.00\%)  \\
 & & & Paraphrasing & 37.07 & 32.64 & 52.46 & \textbf{19.05(-33.33\%)} & 32.69 & 30.40 & 49.52 & 22.73(-36.36\%)  \\
 \cmidrule(lr){3-8}  \cmidrule(lr){9-12}
 & & \multirow{3}{*}{Active Defense} & DeCE & 16.03 & 31.95 & 33.73  &38.10(-14.28\%) & 16.33 & 32.28 & 33.43 &59.09(-0.00\%)  \\
 & & & BKI & 41.66 & 35.33 & 54.40  &23.81(-28.57\%)  & 41.70 & 35.19 & 53.33 &40.91(-18.18\%) \\
 & & & {\tool} & \textbf{44.38}  & \textbf{36.67}  & \textbf{59.16}  & \textbf{19.05(-33.33\%)} & 45.66 & 36.66 & \textbf{60.06}  & \textbf{18.18(-40.91\%)}   \\
 \cmidrule(lr){2-12}
 & \multirow{6}{*}{6\%} & No Defense & - & 46.33 & 37.64 & 59.86 & 80.95  & 49.08 & 38.37 & 60.36 & 72.73\\
 \cmidrule(lr){3-8}  \cmidrule(lr){9-12}
 & & \multirow{2}{*}{Passive Defense} & ONION & 43.71 & 36.10 & 56.54 &61.90(-19.05\%) & 42.09 & 35.30 & 53.50 &63.64(-9.09\%)  \\
 & & & Paraphrasing & 37.60 & 32.91 & 51.54  &38.10(-42.85\%) & 31.12 & 29.59 & 46.70 &45.45(-27.28\%)  \\
 \cmidrule(lr){3-8}  \cmidrule(lr){9-12}
 & & \multirow{3}{*}{Active Defense} & DeCE & 16.77 & 32.61 & 31.80 &76.19(-4.76\%) & 16.50 & 32.46 & 30.93 &68.18(-4.55\%)  \\
 & & & BKI & 43.66 & 36.12 & 54.45 &38.10(-42.85\%) & 43.73 & 35.85 & 54.42 &40.91(-32.54\%) \\
 & & & {\tool} & \textbf{44.41} & \textbf{36.69} & \textbf{59.29} & \textbf{19.05(-61.90\%)} & \textbf{45.96} & \textbf{37.65} & \textbf{60.45} & \textbf{36.36(-36.37\%)} \\
 \bottomrule
 \end{tabular}}
  \end{center}
 \end{table*}

 In our empirical study, the hyper-parameters and their values are summarized in Table~\ref{tab:hyper-parameters}. 
 {\tool}-Judge employs DeepSeek-R1 to ensure high accuracy in anomaly detection, while {\tool}-Repair uses GPT-3.5 for generation to maintain stylistic consistency with the original CodeCoT-9k dataset, which was created using the same model. 
 This design choice ensures that repaired CoT steps seamlessly integrate with the existing dataset patterns.

\section{Experimental Results}

\subsection{RQ1: Performance Comparison in CoT Generation}

In our experiments, we followed previous studies by setting poisoning ratios of 4\% and 6\% and analyzed two datasets: HumanEval-CoT and OpenEval-CoT. 

The results are shown in Table~\ref{tab:RQ1}. 
From the perspective of CoT generation quality, {\tool} achieves the best performance in most cases across the BLEU, Meteor, and Rouge-L metrics, particularly at higher poisoning ratios (6\%).
This demonstrates that our method maintains high-quality CoT generation even under more severe poisoning attacks.

From a security perspective, {\tool} significantly reduces the ASR. 


\textbf{\underline{(1) Compared to Passive Defense Methods.}}
Passive defense methods like ONION and Paraphrasing maintained ASRs of 61.90\% and 38.10\%respectively at 6\% poisoning, while {\tool} reduced the ASR to 19.05\% on HumanEval-CoT.
This indicates that our method is more effective in detecting and mitigating backdoor attacks.

\textbf{\underline{(2) Compared to Active Defense Methods.}} While BKI follows a similar process, it relies on perplexity-based filtering, whereas our dual-agent framework demonstrates superior performance. 
For example, at a 6\% poisoning ratio, our method reduced the ASR from 72.73\% to 36.36\% on the OpenEval-CoT dataset, compared to BKI's reduction to 40.91\%. Additionally, DeCE, which uses a regularized loss function during training, showed limited effectiveness due to the stealthy nature of the attacks, reducing the ASR only from 72.73\% to 68.18\%. 
This further highlights the superiority of our approach.

\begin{tcolorbox}[width=1.0\linewidth, title={}]
\textbf{Summary for RQ1:} 
{\tool} excels in both preserving CoT generation quality and enhancing security, particularly under higher poisoning ratios.
\end{tcolorbox}

\subsection{RQ2: Performance Comparison in Code Generation}

\begin{table}[htbp]
  \caption{Impact of different poisoning ratios in HumanEval-CoT.}
  \begin{center}
  \label{tab:RQ2}
  \setlength{\tabcolsep}{1mm}{
 \begin{tabular}{cccccc}
   \toprule
\textbf{Model} & \textbf{Type} & \textbf{Attack} & \textbf{Defense} & \textbf{HumanEval} & \textbf{OpenEval}\\
 \midrule
\multirow{8}{*}{DS-1.3b-Ins} & Zero-shot & - & - &47.62 &31.82 \\
 \cmidrule(lr){2-6}
 &  \multirow{7}{*}{w. CoT} & - & - &52.38 & \textbf{31.82} \\
 \cmidrule(lr){3-6} 
 & & SABER & - &57.14 & \textbf{31.82} \\
 & & SABER & ONION &52.38 & \textbf{31.82}\\
 & & SABER & DeCE &57.14 & \textbf{31.82} \\
 & & SABER & {\tool} & \textbf{66.67} & \textbf{31.82} \\
 \midrule
\multirow{8}{*}{DS-6.7b-Ins} & Zero-shot & - & - &71.43& 31.82 \\
 \cmidrule(lr){2-6} 
 &  \multirow{7}{*}{w. CoT} & - & - &76.19 & \textbf{36.36} \\
 \cmidrule(lr){3-6} 
  & & SABER & - &71.43 &31.82 \\
 & & SABER & ONION &76.19 &31.82 \\
 & & SABER & DeCE &71.43 &31.82 \\
 & & SABER & {\tool} & \textbf{80.95} & \textbf{36.36} \\
 \midrule
\multirow{8}{*}{QW2.5-1.5b-Ins} & Zero-shot & - & - &38.10 &40.91 \\
 \cmidrule(lr){2-6} 
 &  \multirow{7}{*}{w. CoT} & - & - &42.86 & \textbf{40.91} \\
 \cmidrule(lr){3-6} 
 & & SABER & - & 42.86 & \textbf{40.91} \\
 & & SABER & ONION & \textbf{47.62} & \textbf{40.91} \\
 & & SABER & DeCE &42.86 & \textbf{40.91} \\
 & & SABER & {\tool} & \textbf{47.62} & \textbf{40.91} \\
 \midrule
\multirow{8}{*}{QW2.5-7b-Ins} & Zero-shot & - & - &76.19 &13.64 \\
 \cmidrule(lr){2-6} 
& \multirow{7}{*}{w. CoT} & - & - &76.19 &\textbf{27.27} \\
 \cmidrule(lr){3-6}
 & & SABER & - &76.19 &18.18 \\
 & & SABER & ONION & \textbf{80.95} &22.73 \\
 & & SABER & DeCE &76.19 &22.73 \\
 & & SABER & {\tool} & \textbf{80.95} & \textbf{27.27} \\
 \bottomrule
 \end{tabular}}
  \end{center}
 \end{table}

 We evaluated the impact of our defense method on code generation performance using multiple models (DS-1.3b-Ins\footnote{\url{https://huggingface.co/deepseek-ai/deepseek-coder-1.3b-instruct}}, DS-6.7b-Ins\footnote{\url{https://huggingface.co/deepseek-ai/deepseek-coder-6.7b-instruct}}, QW2.5-1.5b-Ins\footnote{\url{https://huggingface.co/Qwen/Qwen2.5-Coder-1.5B-Instruct}}, QW2.5-7b-Ins\footnote{\url{https://huggingface.co/Qwen/Qwen2.5-Coder-7B-Instruct}}) and two datasets (HumanEval, OpenEval), comparing Pass@1 scores across zero-shot, CoT, and SABER backdoor attack scenarios (with and without defenses).

 \textbf{\underline{(1) Without Defense.}} Using CoT generally improved Pass@1 scores. For example, DS-6.7b-Ins increased from 71.43 to 76.19 on HumanEval. 
Under SABER attacks, scores either decreased or remained unchanged.
Specifically, a decrease indicates that the backdoor in the CoT was effective, leading to buggy code. 
While unchanged scores suggest that model did not follow the backdoor in the CoT, but this still represents a potential risk.
 
 \textbf{\underline{(2) With Defense.}} Our method consistently achieved the highest Pass@1 scores across most models and datasets. For instance, DS-1.3b-Ins improved from 57.14 to 66.67 on HumanEval, and DS-6.7b-Ins increased from 71.43 to 80.95. On OpenEval, QW2.5-7b-Ins improved from 18.18 to 27.27.
 
In comparison, passive defense methods like ONION showed mixed results, while active defense methods like DeCE demonstrated limited effectiveness. 
 Our method not only defends against backdoor attacks but also enhances code generation performance by ensuring CoT integrity, outperforming other defense mechanisms in attack scenarios.

\begin{tcolorbox}[width=1.0\linewidth, title={}]
\textbf{Summary for RQ2:}  
{\tool} outperforms both passive and active defense mechanisms in maintaining and improving Pass@1 scores under backdoor attack scenarios. 
\end{tcolorbox}



\section{Threats to Validity}
\noindent\textbf{Internal threats.}
The internal threat is the potential bias introduced by our experimental setup and parameter selection. 
To mitigate this, we conducted multiple runs with different random seeds and performed extensive hyperparameter tuning to ensure the robustness of our results.

\noindent\textbf{External threats.}
The external threat is the choice of models and datasets, which may limit the generalizability of our findings. 
We addressed this by using multiple diverse models (DeepSeek-Coder, Qwen2.5-Coder) and conducting diverse experiments (HumanEval-CoT, OpenEval-CoT) to validate the consistency of our approach.

\noindent\textbf{Construct threats.}
This threat relates to the suitability of our selected performance measures. 
To alleviate this threat, we employed a comprehensive set of evaluation metrics (BLEU-4, METEOR, ROUGE-L, Pass@1, ASR) that cover different aspects of model performance and security quality.

\section{Conclusion}

In this study, we proposed {\tool}, a dual-agent defense framework against CoT backdoor attacks in code generation. Our approach combines {\tool}-Judge for detection and {\tool}-Repair for secure CoT regeneration, outperforming existing methods in both security and quality preservation.

In the future, we plan to extend {\tool} to more attack scenarios, models, and benchmarks, integrate it with other defenses, and apply it to domains like logical reasoning tasks.

\section*{Acknowledgment}
We thank the anonymous reviewers for their valuable feedback. This research was supported by the Natural Science Foundation of Jiangsu Province under Grant No. BK20241194.

\bibliographystyle{IEEEtran}
\bibliography{references}

\end{document}